\documentstyle[psfig,prb,aps]{revtex}
\draft

\title{Wigner Crystallization of a two dimensional electron gas
in a magnetic field:
single electrons versus electron pairs at the lattice sites}
\author{M. Taut\\Institute for Solid State and Materials Research
Dresden\\ POB 270016,
01171 Dresden, Germany\\
email: m.taut@ifw-dresden.de\\}

\begin{document}
\maketitle

\begin{abstract}
The ground state energy and the lowest excitations of a two dimensional
Wigner crystal in a perpendicular magnetic field with one  and
two electrons per cell is investigated.
In case of two electrons per lattice site,
the interaction of the electrons {\em within} each cell is
taken into account exactly (including exchange and correlation effects),
and the interaction {\em between} the cells is in second order 
(dipole) van der Waals
approximation. No further approximations are made,
in particular Landau level mixing and {\em in}complete spin polarization are 
accounted for. Therefore, our calculation
comprises a, roughly speaking,  complementary description of the 
bubble phase (in the special case of  one and two
electrons per bubble), which was  proposed
by Koulakov, Fogler and Shklovskii 
 on the basis of a Hartree Fock calculation.
The phase diagram shows 
that in GaAs the paired phase is energetically more favorable than the 
single electron phase for, roughly speaking,   filling factor $f$
larger than 0.3 and density parameter $r_s$ smaller than 19 effective 
Bohr radii (for a more precise statement see Fig.s 4 and 5).
If we start within the paired phase and increase magnetic field or 
decrease density, the pairs first undergo some singlet-- triplet 
transitions before they break.
\end{abstract}
\pacs{PACS: 71.10.Hf Non Fermi liquid ground states,\\
            71.10.Li Excited states and pairing interactions in model systems,\\
            73.40.Hm Fractional Quantum Hall Effect }

\section{Introduction}
The Wigner crystal (WC) and the Fermi liquid are states of matter which are 
established already for decades. Later on, the Laughlin liquid as a 
special state in high magnetic fields has been added. In both liquid states, 
wave-functions of the electrons have an appreciable overlap. In the extreme 
Wigner crystal limit this overlap can be neglected. The issue of this paper is 
if there is a state between crystal and liquid, where a finite number of
 particles overlap as in the liquid, but interacts with the neighboring 
cluster as in the Wigner crystal (paired or clustered Wigner crystal). 
The charge density wave (CDW) state, mostly investigated
in Hartree Fock (HF) approximation, is a alternative intermediate phase.
\cite{Maki-Zotos,Yoshioka-Lee,MacDonald-Cote}
In the latter case, 
there is a density modulation, but the density is so high that overlap 
plays an important role. 
(Observe that in some papers the CDW state is also called WC.) 
What we call here 'paired Wigner crystal' can be viewed as a derivative of
three other states. It is as (i) a Wigner crystal \cite{Maradudin} 
with two electrons per cell, 
(ii) a crystallized paired electron liquid known from superconductivity and
theories for the $f=\frac{5}{2}$ state 
\cite{Halperin,Greiter,Moore-Read,Morf,Haldane-paired}
, (iii) 
a state of liquid islands, if the clusters are large. \\

The first paper which considered the 
possibility of a paired WC in 3 dimensions (3D) without a magnetic field is 
Ref.\onlinecite{Ashcroft} using a cell approximation for the inter-cell 
interaction and a variational ansatz for the intra-cell problem.
In a previous paper \cite{Taut-WC3d} it was shown, that the model of Ref.
\onlinecite{Ashcroft} can
 be solved exactly, even if fluctuation corrections in second order 
van der Waals approximation are included additionally. It turned out that in 3D 
without a magnetic field, the paired WC is energetically lower than the 
conventional WC for $r_s<9$, but in this density region the Fermi liquid is
probably already lower than both WCs.
More recently, this issue has been investigated in 2D with a magnetic field
 using the HF approximation.\cite{Shklovskii}
 The result is that in higher 
Landau levels (LL) the ground states are phases with circular or stripy 
clusters. 
Meanwhile, numerical solutions for finite size systems have been published 
\cite{Haldane,Yoshioka}, 
which confirm these findings qualitatively within their model. 
All previous work on this issue neglect 
LL mixing and incomplete spin polarization.
In the present approach, we investigated clusters with one 
and two electrons.
In the latter case,
the interaction within each cluster
is  exactly accounted for (including exchange and correlations). In both cases,
the inter-cell interaction is considered in second order van der Waals
approximation (no exchange, Coulomb interaction in dipole approximation).
In Sect. II we describe the solution of the Schr\"odinger equation in 
more detail and in Sect. III our results are presented. 
Sect. IV gives a summary and an overview.

\newpage

\section{Solution of the Schr\"odinger Equation }
The Hamiltonian of a system of $N$ electrons at each lattice site
 and a positive compensating background of density $n_0$ reads
(atomic units $\hbar=m=e=1$ are adopted)
\begin{equation}
H=T+V_{intra}^{ee}+V_{inter}^{ee}+V^{eb}+V^{bb}
\end{equation}
with the kinetic energy
\begin{equation}
T=\sum_n\sum_k \frac{1}{2m^*}
\left[ {\bf p}_{nk}+ {1\over c} {\bf A}({\bf u}_{nk}) \right]^2
\end{equation}
the intra-cell e--e--interaction
\begin{equation}
V_{intra}^{ee}=\sum_n \frac{1}{2} \sum_{k \neq k'}
\frac{\beta}{|{\bf u}_{nk}-{\bf u}_{nk'}|}
\end{equation}
the inter-cell e--e--interaction
\begin{equation}
V_{inter}^{ee}=\frac{1}{2} \sum_{n \ne n'} \sum_{kk'}
\frac{\beta}{|({\bf R}_n^0-{\bf R}_{n'}^0)+
({\bf u}_{nk}-{\bf u}_{n'k'})|}
\end{equation}
the electron-- background interaction
\begin{equation}
V^{eb}=-\sum_n\sum_k\int d{\bf r}' \frac{n_0}{|{\bf r}_{nk}-{\bf r}'|}
\end{equation}
and the background-- background interaction
\begin{equation}
V^{bb}=\frac{1}{2}\int d{\bf r} \int d{\bf r}' 
\frac{n_0 n_0}{|{\bf r}-{\bf r}'|}
\end{equation}
In these definitions, $m^*$ is the effective mass, $\beta$ the inverse 
background dielectric constant, ${\bf R}_n^0$ are the lattice sites,
and ${\bf r}_{nk}={\bf R}_n^0+{\bf u}_{nk}$ the electron coordinates.
We use the symmetric gauge ${\bf A}=\frac{1}{2}\;{\bf B}\times {\bf u}$. 
Latin letters $n$ and $n'$ denote the lattice sites and $k$ and $k'$
the electrons within a cell (or cluster).
If the distance between the cell centers is large compared with the average
 distance between the electrons within a cell, the inter-cell interaction can
be expanded in a multi-pole series up to second order
\begin{equation}
V_{inter}^{ee}=V_{inter}^{ee (0)}+V_{inter}^{ee (2)}
\end{equation}
with the zero and second order terms
\begin{eqnarray}
V_{inter}^{ee (0)}&=&N^2 \frac{1}{2} \sum_{n \ne n'} 
\frac{\beta}{|{\bf R}_n^0-{\bf R}_{n'}^0|}\\
V_{inter}^{ee (2)}&=& \frac{1}{N} \frac{1}{2}
\sum_{n n'} \sum_{k k'}
{\bf u}_{n k} \cdot {\bf C}_{n k,n' k'} \cdot 
{\bf u}_{n' k'}
\end{eqnarray}
the force constant matrix
\begin{eqnarray}
{\bf C}_{n k,n' k'}&=&+\beta N^2 {\bf T}_0 
\;\;\;\; for \;\;\; (n k)=(n' k')\\
&=&-\beta N {\bf T}_{nn'}\;\;\; else
\end{eqnarray}
the dipole tensor
\begin{equation}
{\bf T}_{nn'}=\frac{1}{R_{nn'}^5} \left[ 3 \; {\bf R}_{nn'} \circ 
{\bf R}_{nn'}- R_{nn'}^2 \;{\bf I} \right]
\end{equation}
and the definitions
${\bf T}_0=\sum_{n\neq0} \; {\bf T}_{0n}$ 
and $R_{nn'}={\bf R}_n^0-{\bf R}_{n'}^0$.
By rearrangement of terms we obtain the following form
\begin{equation}
H=E_N^{Mad}+(T+V_{intra}^{ee})+ V_{inter}^{ee (2)}
\label{H-harmon}
\end{equation}
with the Madelung energy for a lattice with $N$ electrons at each lattice
site
\begin{equation}
E_N^{Mad}= N^2 \frac{1}{2} \sum_{n \ne n'} 
\frac{\beta}{|{\bf R}_n^0-{\bf R}_{n'}^0|}
-N\sum_n\int d{\bf r}' \frac{n_0}{|{\bf R}_n^0-{\bf r}'|}
+\frac{1}{2}\int d{\bf r} \int d{\bf r}' 
\frac{n_0 n_0}{|{\bf r}-{\bf r}'|}
\label{E-Madelung}
\end{equation}

By introducing the center of mass (c.m.) coordinate ${\bf U}_n$ and some 
relative coordinates within each cell using a
orthogonal coordinate transformation, the last 3 terms of
(\ref{H-harmon}) can be decomposed into a c.m. Hamiltonian $H_{c.m.}$ and
a lattice sum of internal (relative coordinate)
 Hamiltonians $H_{rel}=\sum_n H_{rel,n}$, where 
$H_{rel,n}$ contains only the $(N-1)$ relative coordinates in cell $n$.
Thus, only $H_{c.m.}$ shows a coupling between the cells.
This fact is completely analogous to the treatment of interacting 
quantum dot lattices considered in Ref.\onlinecite{Qdot-lattice}.
Consequently, the total Hamiltonian (\ref{H-harmon}) decomposes into 3 
contributions
\begin{equation}
H=E_N^{Mad}+H_{cm}+H_{rel}
\end{equation}
which are considered in turn.

\subsection{Madelung energy}
Generally, we obtain $E_N^{Mad}$ from the conventional Madelung energy 
($N=1$) with the same lattice constant
by $E_N^{Mad}=N^2 E_{N=1}^{Mad}$. 
Here, however, we compare phases with the same mean density.
In the {\em hexagonal} lattice, 
the Madelung energy per electron $\varepsilon_N^{Mad}$ for a 
given density parameter $r_s=1/\sqrt{\pi n_0}$ can be deduced from the data in
 Ref.\onlinecite{Maradudin} providing
$\varepsilon_N^{Mad}=-1.106103 \; \beta \; N^{1/2} \; r_s^{-1}$.\\

\subsection{Center--of--mass energy}

The c.m. part of the Hamiltonian reads
\begin{eqnarray}
H_{cm}&=&\frac{1}{N}
\bigg\{ \sum_n
\frac{1}{2 m^*}
\left[
{\bf P}_n+
\frac{N}{c} {\bf A}\left({\bf U}_n \right)
\right]^2  \nonumber \\
 && + \frac{N^2}{2} \sum_{n,n'}\; {\bf U}_n \cdot
{\bf C}_{n,n'}
\cdot {\bf U}_{n'}
\bigg\}
\label{H-cm}
\end{eqnarray}
with 
\begin{eqnarray}
{\bf C}_{n,n'}&=&+\beta N^2 {\bf T}_0 
\;\;\;\; for \;\;\; n=n'\\
&=&-\beta N {\bf T}_{nn'}\;\;\; else
\end{eqnarray}
After the usual phonon transformation
\begin{eqnarray}
{\bf U}_n&=&\frac{1}{\sqrt{N_c}} \sum_{\bf q}^{BZ}
e^{-i{\bf q}\cdot R_n^0}\;{\bf U}_{\bf q}\\
{\bf P}_n&=&\frac{1}{\sqrt{N_c}} \sum_{\bf q}^{BZ}
e^{+i{\bf q}\cdot R_n^0}\;{\bf P}_{\bf q}
\label{phonon-trafo}
\end{eqnarray}
where $N_c$ is the number of cells, the Hamiltonian decouples with 
respect to the lattice sums and we obtain 
\begin{equation}
H_{cm}= \sum_{\bf q}^{BZ}\;H_{\bf q}
\end{equation}
with the magneto-phonon Hamiltonian
\begin{eqnarray}
H_{\bf q}&=& \frac{1}{N} \bigg\{   \frac{1}{2 m^*}
\left[
{\bf P}_{\bf q}+\frac{N}{c}{\bf A}({\bf U}_{\bf q}^*)
\right]^\dagger \cdot
\left[
{\bf P}_{\bf q}+\frac{N}{c}{\bf A}({\bf U}_{\bf q}^*)
\right] \nonumber\\
& &+ \frac{N^2}{2} 
{\bf U}_{\bf q}^* \cdot
{\bf C}_{\bf q}
\cdot {\bf U}_{\bf q}  \bigg\}
\label{H-phonon}
\end{eqnarray}
and the dynamical matrix
\begin{equation}
{\bf C}_{\bf q}=\sum_n \;
e^{i{\bf q}\cdot {\bf R}_n^0}\;
{\bf C}_{n,0}=\beta \; N \; \sum_{n \neq 0} \; 
\left( 1-e^{i{\bf q}\cdot {\bf R}_n^0} \right) \; {\bf T}_{n0}
\label{dyn-mat}
\end{equation}
The eigenvalues of $H_{cm}$ and $H_{\bf q}$ 
depend on $N$ only through the $N$ dependence of ${\bf C}$. The explicit
$N$ dependence in (\ref{H-cm}) and (\ref{H-phonon}) cancels. The last 
statement is obvious if we consider that the explicit $N$ can be removed 
by rescaling the displacement vector $\bf U$ by a factor of $\sqrt{N}$.

The eigenvalues of (\ref{H-phonon}) are e.g. given in  
Ref.\onlinecite{Qdot-lattice}. 
\begin{equation}
E(n_+,n_-)= (n_+ +\frac{1}{2})\; \omega_+ + (n_- +\frac{1}{2})\; \omega_- ~;
~~~n_\pm=0,1,2,...
\label{E-N=1}
\end{equation}
This provides a ground state energy 
\begin{equation}
E_{cm}=\sum_{\bf q}^{BZ} \; \frac{1}{2} (\omega_+ + \omega_-)
\label{E-cm}
\end{equation}
where
\begin{equation}
\omega_\pm = \sqrt{
\frac{\omega_c^{*2}}{2}+\tilde\omega_0^2 \pm
\sqrt{\frac{\omega_c^{*4}}{4}+\omega_c^{*2}\;\tilde\omega_0^2+\frac{\Delta^2}{4}
+C_{12}^2}}
\label{omega-pm}
\end{equation}
with
$\tilde\omega_0^2=\frac{1}{2}(C_{11}+C_{22})$, 
$\Delta=C_{11}-C_{22}$, 
$\omega_c^*=\frac{B}{m^* c}$ is the cyclotron frequency
with the effective mass, and $C_{i,k}$ are the Cartesian components
of ${\bf C}_{\bf q}$.
If we define an auxiliary tensor $\bf S$ through
 ${\bf C}_{\bf q}=p\;{\bf S}_{\bf q}$ with
$p=2\beta N/a^3$ ($a$ is the lattice constant), then ${\bf S}_{\bf q}$ 
depends only from the geometry of the lattice, i.e. it is a fixed tensor
for the hexagonal and cubic lattice and depends only from the $b/a$ ratio 
for a rectangular lattice.
The explicit form of ${\bf S}_{\bf q}$ is complicated and not of common
interest. The sums involved have to be done numerically anyway.
The interaction parameter $p$ for the hexagonal lattice reads in terms of the 
more convenient parameters $r_s$ and $n_0$
\begin{equation}
p=2\left( \frac{\sqrt{3}}{2\pi}\right)^{3/2} \beta \; r_s^{-3} N^{-1/2}=
 2\left( \frac{\sqrt{3}}{2}\right)^{3/2} \beta \; n_0^{3/2} N^{-1/2}
\label{p}
\end{equation}
Because of the optical selection rules, $\omega_{\pm}$ for ${\bf q}=0$ 
agree with the
excitation energies $\Delta E$ for long wavelength (infrared) radiation.\\
It can be  easily seen that for $B=0$, 
$\omega_\pm=\sqrt{eigenvalues({\bf C}_{\bf q})}=
\sqrt{p}\;\sqrt{eigenvalues({\bf S}_{\bf q})}$. 
Because of the universal character of ${\bf S}_{\bf q}$, 
the excitation energies in units of $\sqrt{p}$ comprise 
all data on the energies for
hexagonal lattices in this limit (see Fig.1).
In connection with formula (\ref{p})
this means that the phonon frequencies
  of  the paired WC, as
compared with the single electron case of the same electron density,
are smaller by a factor of $1/\sqrt{2}$.
Additionally, the frequencies (and the c.m. contribution to the total 
energy) for the same $N$ decay with the density
parameter like $r_s^{-3/2}$.

\begin{figure}[th]
\vspace{-2cm}
\begin{center}
{\psfig{figure=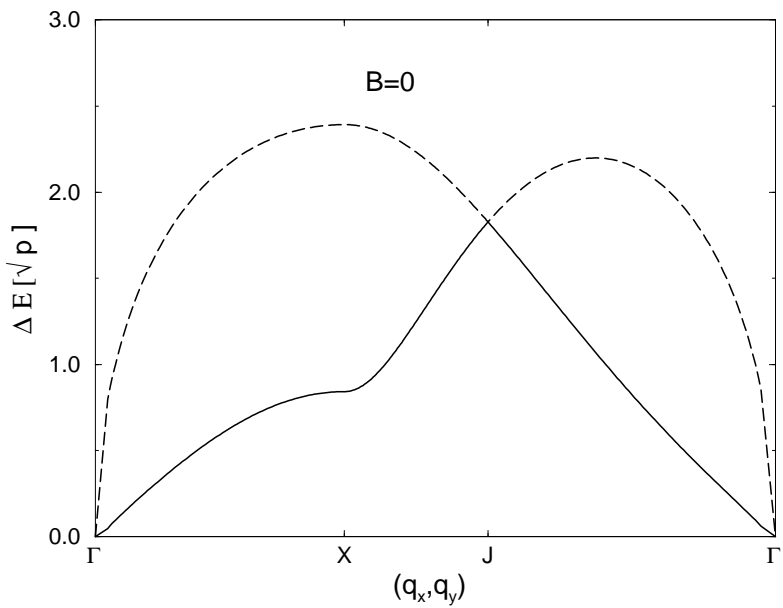,angle=-90,width=12.cm,bbllx=15pt,bblly=45pt,bburx=580pt,bbury=750pt}}
\caption[]{Phonon dispersion (for B=0) of the hexagonal Wigner lattice.
(Energies are in units $\sqrt{p}$.)
}
\label{Fig1}
\end{center}
\end{figure}

For finite $B$, an universal function
comprising all data cannot be defined. However, if we use $\omega_c^*$
as energy unit and $\omega_c^{*2}$ as unit for the interaction parameter $p$,
then the result depends only on $p$ and shows {\em no explicit}
dependence on $B$.\\
For large $B$, (\ref{omega-pm}) gives
$\omega_+=\omega_c^* + trace({\bf C}_{\bf q})/2 \omega_c^* $
and $\omega_-=determinant({\bf C}_{\bf q})/\omega_c^*$ 
(in agreement with Ref.\onlinecite{MacDonald-Cote}),
and in the units mentioned above, this reads
$\omega_+[\omega_c^*]=1+p[(\omega_c^*)^2]\cdot trace({\bf S}_{\bf q})/2$
and $\omega_-[\omega_c^*]=p[(\omega_c^*)^2]\cdot determinant({\bf S}_{\bf q})$.
Thus, the results are determined by the interaction parameter and 
two universal functions, namely
 $trace({\bf S})$
and $determinant({\bf S})$, which are shown in  Fig.2.
In the limit $B\rightarrow \infty$ 
we obtain the result of noninteracting electrons
  $\omega_+=\omega_c^*$ and $\omega_-=0$, as to be expected. 

\begin{figure}[th]
\vspace{-2cm}
\begin{center}
{\psfig{figure=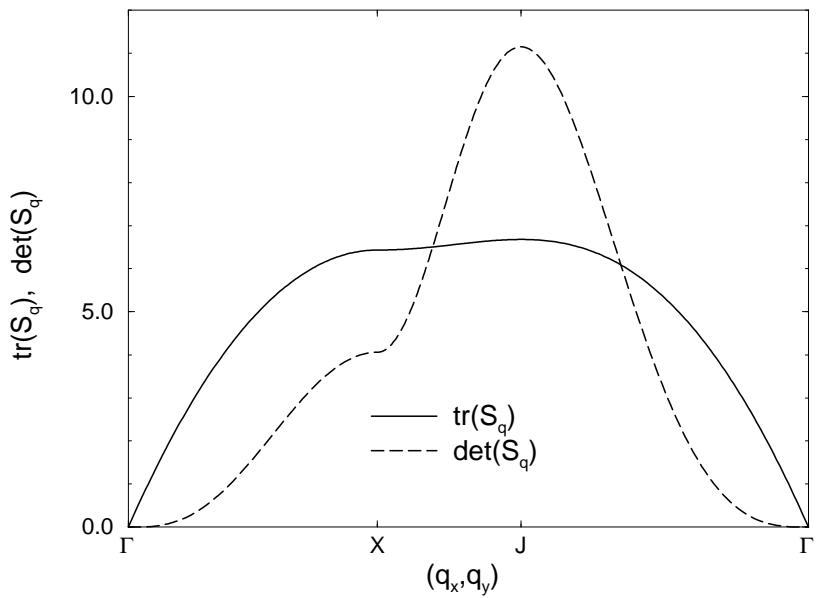,angle=-90,width=12.cm,bbllx=15pt,bblly=45pt,bburx=580pt,bbury=750pt}}
\caption[]{ The functions $trace({\bf S}_{\bf q})$ and $determinant({\bf S}_{\bf q})$ determining  the excitation frequencies in the high field limit.
}
\label{Fig2}
\end{center}
\end{figure}

Direct comparison of excitation energies with experiments is not the aim of
this work. What matters here is only the contribution of the  'zero
point vibrations' to the total energy.

\subsection{Internal energy}
For two electrons per cell, the internal (relative coordinate)
 energy can be calculated easily.
 (Larger N call for a larger numerical effort and are postponed to a 
succeeding work.)
If we introduce apart from the c.m. ${\bf U}_n$ 
the relative coordinate ${\bf u}_n$
\begin{eqnarray}
{\bf U}_n&=&\frac{1}{2}({\bf u}_{n1}+{\bf u}_{n2})\\
{\bf u}_n&=&{\bf u}_{n2}-{\bf u}_{n1}
\end{eqnarray}
then all relative Hamiltonians read (the index 'n' is omitted)
\begin{equation}
H_{rel}=2 \bigg\{
\frac{1}{2 m^*}
\left[
{\bf p}+
{1\over 2 c} {\bf A}({\bf u})
\right]^2 +
\frac{1}{2}\; {\bf u} \cdot {\bf D } \cdot {\bf u}
+\frac{\beta}{2\;u}
\bigg\}
\label{H-rel}
\end{equation}
where
${\bf p}=-i \nabla_{\bf u}$ and
${\bf D}=\frac{\beta}{2} {\bf T}_0=\frac{1}{4} p S_0 {\bf I}$ with 
$S_0$ being a lattice sum of the same type as the components of $S_{ik}$.
(Do not mix up the interaction parameter $p$ and the {\em vector}
of the  momentum operator $\bf p$.) It should be emphasized that the 
electronic
{\em inter}cell interaction contributes a harmonic term 
to the {\em intra}cell problem to $H_{rel}$, namely 
the last but one in (ref{H-rel}).
As to the calculation of the eigenvalues of Hamiltonians of the form 
(\ref{H-rel}) we refer to 
Ref.\onlinecite{Taut2e-in-B}.

\newpage
\section{Results}
If not otherwise indicated, all results and parameters are given in
effective atomic units for GaAs ($m^*=0.067, \beta=1/12$),.i.e.,
energies (and frequencies)  in $1 a.u.^*=4.65\cdot 10^{-4}\; double Rydberg=
12.64 \; meV$ and length in $1 a.u.^*= 1.791\cdot 10^{2} Bohr=
0.9477 \cdot 10^{2} \AA$. (If the units of a quantity is
indicated explicitly, then it is given in brackets after the quantity.) 
Observe that the effective density parameter
in current typical experiments is of order 2.

\subsection{Contributions to the total energy}
For larger $r_s$, the energy difference between both phases is tiny (see
Fig.3).
As an example, at $r_s=10$ (at the right boundary of Fig.3)
it amounts to $2\cdot10^{-4}$,
 calling for
high precision computations (of order $10^{-5} a.u.^*$).
This fact is amazing because the Madelung energy and the c.m. energy are both
very different (see Fig.3). This difference is compensated almost completely
by the relative energy of the paired phase.

\begin{figure}[th]
\vspace{-2cm}
\begin{center}
{\psfig{figure=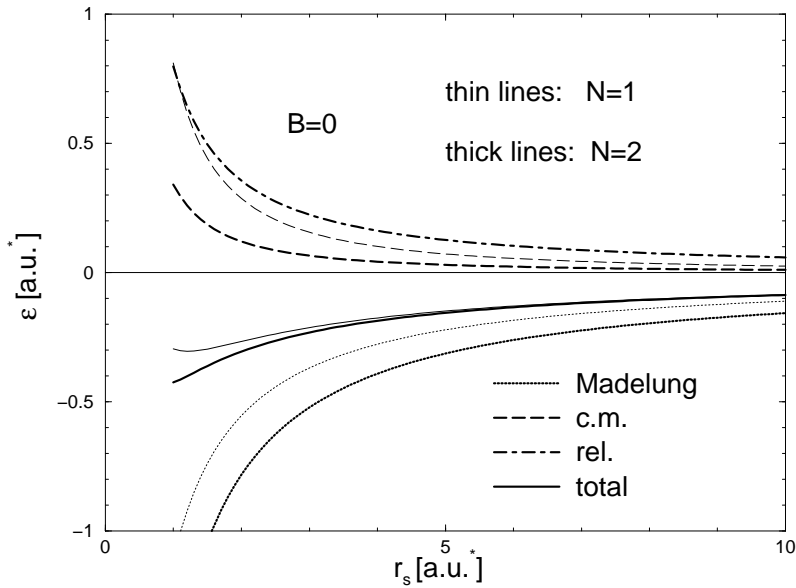,angle=-90,width=12.cm,bbllx=15pt,bblly=45pt,bburx=580pt,bbury=750pt}}
\caption[]{ Comparison of the contributions to the total energy in the
single electron (N=1) and the paired (N=2) Wigner crystal for zero
magnetic field.
}
\label{Fig3}
\end{center}
\end{figure}

\subsection{Phase diagram for single electron versus paired Wigner crystals}
 Fig.4 shows the phase diagram for the single electron versus the paired
 WC phase
in the $r_s$-- $B$ plane. In shaded regions, the electron pairs are in singlet 
state and in triplet states elsewhere. The spin configuration of the 
paired phase has been 
determined irrespective of the question, if the single electron phase 
is below both paired (singlet or triplet) phases. The thin full lines separate 
regions with different  relative (internal) angular 
momentum $m$ of the pairs, 
which has the lowest energy. The value of $m$ is indicated in 
each region as well. 
(Even $m$ belongs to the state and odd $m$ to the triplet state.)
The thick full line separates the regions where the single electron WC 
is the ground state from regions, where the paired WC has the lower energy.
We want to emphasize that this figure does not say anything about the issue 
if  any other (e.g. liquid) phase is lower in energy than the
phases considered here. 
The kinks in the phase boundary  are connected with
changes of the total spin of the paired phase.

\begin{figure}[th]
\vspace{-2cm}
\begin{center}
{\psfig{figure=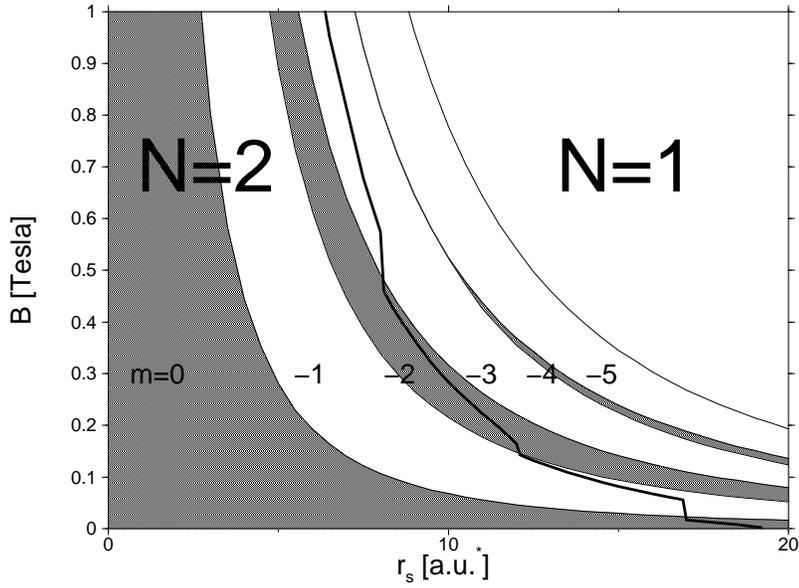,angle=-90,width=12.cm,bbllx=15pt,bblly=45pt,bburx=580pt,bbury=750pt}}
{\psfig{figure=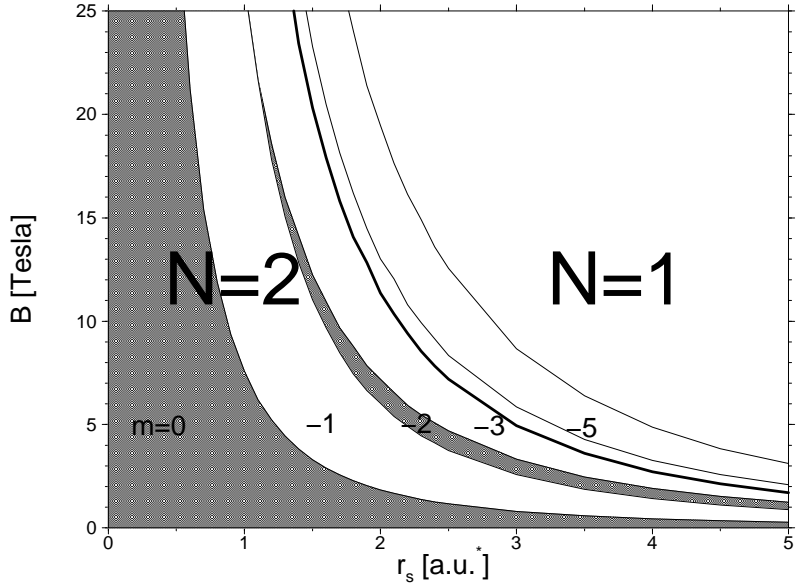,angle=-90,width=12.cm,bbllx=15pt,bblly=45pt,bburx=580pt,bbury=750pt}}
\caption[]{ Phase diagram in the $r_s$-- $B$ plane (see text for details) a) for
small $B$ and b) for small $r_s$.
}
\label{Fig4}
\end{center}
\end{figure}

First, we discuss the case $B=0$. For $r_s$ above 19,
 the energetically
lower state is the
single electron WC and below it is the paired phase.
Clearly, in the limit of high densities (small $r_s$)
any liquid state should win.
It is physically imaginable, that
in increasing the density, the state starts with the single electron WC, then
adopts the paired state before it goes over to the liquid. If there  are
states with 3, 4, ... electrons involved, will be the focus of further
investigations.\\
For a certain finite $B$, the density, where the transition to the
paired ground state occurs, becomes higher. In other words, for higher $B$ one
needs less dilute electron systems to produce a single electron WC.
This is physically obvious, because the magnetic field helps to localize
 the electrons. In other words, lowering the density cases pair breaking.
Now we discuss the pair breaking process as a function of $B$ for fixed $r_s$, 
say $r_s=2$. If we start with $B=0$ and increase $B$, the paired state 
undergoes first some spin transitions from singlet to triplet and
 vice versa before the pairs break in the triplet state. It should be
mentioned that
pair breaking and spin transition are not coupled together.
The reason is that the Zeeman energy is too small to play a decisive role.

\begin{figure}[th]
\vspace{-2cm}
\begin{center}
{\psfig{figure=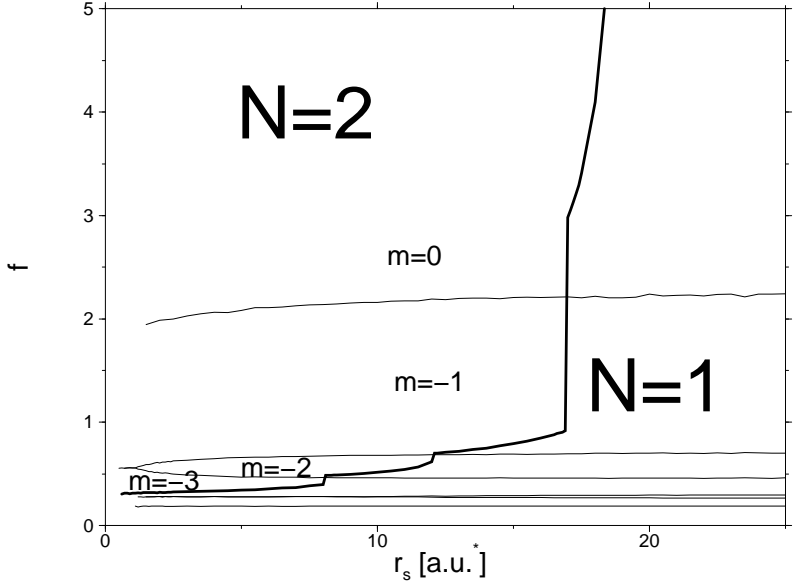,angle=-90,width=12.cm,bbllx=15pt,bblly=45pt,bburx=580pt,bbury=750pt}}
\caption[]{ Phase diagram in the $r_s$-- $f$ plane (see text for details).
}
\label{Fig5}
\end{center}
\end{figure}

Fig.5 shows the the phase diagram again, but the magnetic field axis is
replaced by the filling factor  $f=14.64/(B[T] \; r_s^2[a.u.^*])$ (using the
parameters of GaAs). 
This plot has the advantage of showing the physically important parameter
$f$ directly, 
but does not allow the discussion of the limit $B=0$, because it is transformed
 to $f\rightarrow \infty$.)
Again, thin lines separate regions with different 
$m$ (and therefore total spin) of the paired phase and the thick line
separates the single electron WC ground state from the paired WC.
As to be expected, the single electron WC is restricted to the low density--
small filling factor region. It is clear that in a complete phase diagram, 
the region, which is here
attributed to the paired phase, can be further 
restricted by the 
inclusion of all kinds of liquid phases
 (and possibly WC phases with N=2,3,... electrons per cell). 

\subsection{Comparison of Wigner crystal phases with Laughlin liquid}
Because the Laughlin liquid is restricted to discrete filling factors, 
we compare in Fig.6 the energy (or, more conveniently, 
$r_s (\varepsilon-\omega_c/2)$)
of different phases
for fixed $f=1/3$ as a function of $r_s$. 
The Zeeman energy is omitted in this figure 
because the paired phase is triplet over the
whole $r_s$ range. Therefore it would have contributed for a given $r_s$ 
the same amount to all
phases, but introduce an unpleasant pole at $r_s=0$ into the curves.
(In all calculations, however, the Zeeman energy has to be taken into account 
because of the different spin configurations in the paired WC.)
The curve for the Laughlin liquid from Ref.\onlinecite{Price} includes 
Landau level mixing (as our results
do) by adding a variational correlation factor to the Laughlin function. 
(The value at $r_s=0$ corresponds to the result of the original 
Laughlin function.)
As seen in Fig.6, the Laughlin liquid and the single electron WC are the 
ground state for $r_s$ below and above 15, respectively.
This is in qualitative agreement with experiment.
In our result for $f=1/5$ (not shown), the single electron WC 
is the ground state over the whole 
range of densities. This is not quite correct, because experiments suggest that
the domain of the WC begins at $f=1/7$. Inclusion of a variational
 correlation factor to the WC wave-function fixes  the phase boundary
between the Laughlin liquid and the WC
\cite{Louie,Lam-Girvin}. This amendment could  not be added here, because 
it would have become too complicated for the paired WC and 
in comparing both WC phases we 
needed a description on the same footing.

\begin{figure}[th]
\vspace{-2cm}
\begin{center}
{\psfig{figure=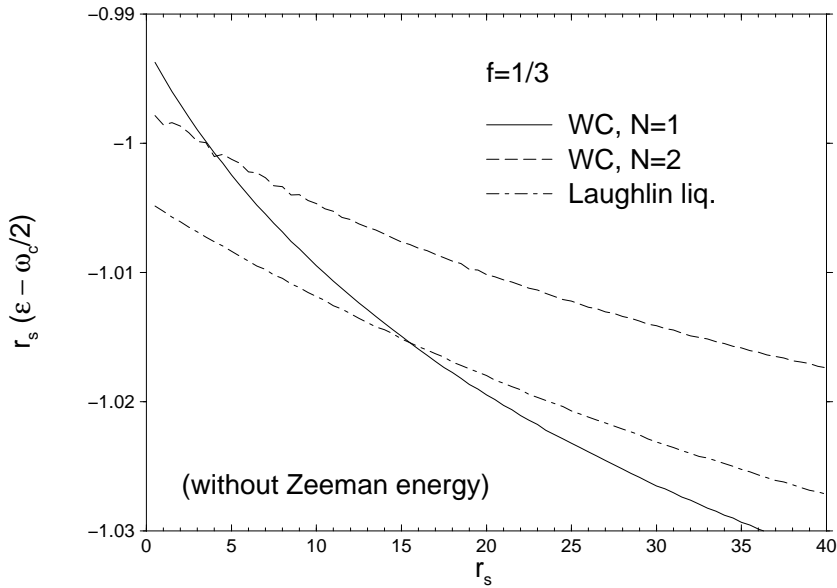,angle=-90,width=12.cm,bbllx=15pt,bblly=45pt,bburx=580pt,bbury=750pt}}
\caption[]{Comparison of the Wigner crystal states with the Laughlin liquid.
}
\label{Fig6}
\end{center}
\end{figure}

\section{Summary and overview}
We have shown that the paired WC is energetically more 
favorable than the single electron WC for higher densities and lower
 magnetic fields. If we start with the paired WC, lowering the density and 
increasing the magnetic field will break the pairs.
At least in GaAs, the Zeeman energy plays a less important role than the
angular momentum dependence of the internal (relative coordinate) energy.
Observe that changing the internal angular momentum (from even to odd) 
is coupled with a change of the total spin of each cluster 
due to the Pauli principle.
Therefore, the paired phase undergoes first some singlet-- triplet transitions 
before the pairs break and the single electron WC is reached. 
The singlet-- triplet transitions have the same physical origin as the 
corresponding transitions in quantum dots, because they are driven by the
internal energy, which is similar in either system.\\
In our approach,
the interaction within each cluster 
is  exactly accounted for (including exchange and correlations). In both cases,
the {\em inter}cell interaction is considered in second order van der Waals 
approximation (no exchange, Coulomb interaction in dipole approximation). 
The validity of the second order approximation has been checked
in Ref.\onlinecite{Esfarjani-Chui} by estimating the third order contribution,
 which amounts to a few \%. We did not include higher order contributions 
from two reasons. First, they spoil the exact decoupling 
and make the paired WC intractable. Second, these
 corrections have most likely the same sign in the single electron 
and the paired phase 
and they cancel partly in comparison of both energies.
The neglect of inter-cell exchange (but keeping the Coulomb correlations) 
is justified in the low density and 
high magnetic field limit, because the exchange energy drops off 
exponentially, but the correlation energy like $1/r$.
It seems to be worthwhile to elaborate on the differences of this work and 
Ref.\onlinecite{Shklovskii}. They targeted on high LLs (low magnetic fields), 
and high electron densities ($N_{LL}>>r_s^{-1}>>1$). LL mixing and incomplete
spin polarization are ruled out and the completely filled LLs 
are frozen out and influence the 
upper LL only through a screening in the e-- e-- interaction.
Therefore we called both treatments in the abstract 'complementary'.
What seems to be interesting is that despite approaching reality 
with different assumptions and virtually 
from opposite sides, the effect of clusterization of the liquids is 
found in both treatments.

\end{document}